\documentclass{article}

\usepackage{amssymb, amsmath, esdiff, esint, upgreek}
\usepackage{graphicx}
\usepackage[colorlinks, urlcolor=blue, unicode]{hyperref}
\usepackage[
	style = phys,
	biblabel = brackets,
	eprint = true,
	bibencoding=auto,
	backend=biber,
	sorting=none,
	language=english,
	autolang=other,
	doi=false,
	url=false,
	maxbibnames=3
	]{biblatex}

\usepackage{authblk}

\AtEveryBibitem{
	\clearfield{day}
	\clearfield{month}
	\clearfield{endday}
	\clearfield{endmonth}
}
\AtEveryBibitem{\clearfield{issn}}
\AtEveryCitekey{\clearfield{issn}}

\bibliography{references}

\title{On mass spectra of primordial black holes}

\author[1]{
    Alexander A. Kirillov\thanks{\href{AAKirillov@mephi.ru}{AAKirillov@mephi.ru}} 
    }
\author[1,2]{
    Sergey G. Rubin\thanks{\href{SGRubin@mephi.ru}{SGRubin@mephi.ru}}
    }

\affil[1]{
    National Research Nuclear University MEPhI
    \par
    (Moscow Engineering Physics Institute), 
    \par
    115409 Kashirskoe shosse 31, Moscow, Russia
    }
\affil[2]{
    N. I. Lobachevsky Institute of Mathematics and Mechanics,
    \par
    Kazan Federal University, 
    \par
    420008 Kremlevskaya street 18, Kazan, Russia
    }

\date{}

\newcommand{\D}{\mathrm{d}\,\!}
\newcommand{\Chi}{\ensuremath\mathrm{X}}

\DeclareMathOperator{\const}{const}

\begin{document}

\maketitle

\begin{abstract}
    Evidences for the primordial black holes (PBH) presence in the early Universe renew permanently. New limits on their mass spectrum challenge existing models of PBH formation. One of the known model is based on the closed walls collapse after the inflationary epoch. Its intrinsic feature is multiple production of small mass PBH which might contradict observations in the nearest future.
    % We show that the mechanism of walls collapse can be applied to produce substantially different PBH mass spectra. 
    % Necessary element of our study is the accounting of the scalar field classical motion together with its quantum fluctuations at the inflationary stage.
    We show that the mechanism of walls collapse can be applied to produce substantially different PBH mass spectra if one takes into account the classical motion of scalar fields together with their quantum fluctuations at the inflationary stage.
\end{abstract}

\section{Introduction}

Interest in the primordial black holes (PBHs) is dramatically increasing since the gravitational waves discovery from the black holes mergers \cite{2016PhRvL.116f1102A}. However, PBHs origin and possible formation mechanisms are still a topical issue of the modern astrophysics and cosmology. The first ideas of such mechanisms had been proposed in \cite{1967SvA....10..602Z, 1971MNRAS.152...75H, 1974MNRAS.168..399C} and lately developed in many other works (see reviews and references within \cite{2010RAA....10..495K, 2020ARNPS..7050520C}). The different PBH spectra are used in papers \cite{1975ApJ...201....1C, 1993PhRvD..47.4244D, 2006JCAP...06..003S, 2015PhRvD..92b3524C, 2016JCAP...02..064G, 2019PhRvD..99j3535C, 2020PhRvD.101b3513L} depending on specific needs.

The phase transitions of the first \cite{1982PThPh..68.1979K, 1982PhRvD..26.2681H, 1998SvPAJ..24....1K, 1999PhRvD..59l4014J, 1999PAN....62.1593K} and the second order \cite{2000hep.ph....5271R, 2001JETP...92..921R} might also underlay a mechanism of the PBH formation.
In this paper, we continue elaboration of the model based on the second type phase transitions during the inflationary epoch \cite{2000hep.ph....5271R, 2001JETP...92..921R, 2002astro.ph..2505K}. However, the described model has a flaw. It inevitably leads to a multiple production of small mass PBHs. That problem could not be avoided within the framework of the discussed scenario, and typical mass spectra have the falling form $\D{N}/\D{M} \propto M^{-\alpha}$, $\alpha>0$ (see review \cite{2019EPJC...79..246B}). Such form of spectra could be unfavorable for explaining the observable effects. Moreover, the overproduction of low-mass PBHs could contradict future experiments. The way out discussed in this paper consists of an involving the classical motion of scalar fields together with their quantum fluctuations.

We study the way to circumvent the problem supposing a complicated form of scalar field potential. The latter is used in a variety of inflationary models predicting the potential landscape. Moreover, the inflaton might consist of multiple fields \cite{2007LNP...738..275W}. For instance, the supergravity often produces more than one physical scalar field \cite{2012JCAP...08..022K} and predicts nontrivial forms of inflaton potentials \cite{2021Univ....7..115K}. The string theory also predicts the landscape with a large number of vacua, local peaks and saddle points \cite{2003dmci.confE..26S, 2005hep.th....1179C}. Such complicated potential can be presented both random and quasi-periodic shape \cite{2009PhRvD..80l3535L} and leads to the multi-field inflation such as multi-stream inflation \cite{2009JCAP...07..033L, 2012JCAP...04..012D}, assisted inflation \cite{2009JCAP...03..027B} or multi-field inflation with a random potential \cite{2009JCAP...04..018T, 2012JCAP...02..039F}. Therefore, other non-inflaton scalar fields might have complicated potential as well.

This paper is organized as follows. 
In Section \ref{sec:ClassQuantMot}, we elaborate the way to involve the classical part of scalar fields into the expression for its fluctuations probability. 
The PBH spectrum depends on an initial position of the scalar field that allow us to adjust the model predictions to future observational data without inserting small parameters. The numerical results are represented in Section \ref{sec:PBHForm}. Finally, Section \ref{sec:Concl} concludes the paper.

\section{Quantum fluctuations accompanied by classical motion at the inflationary stage}
\label{sec:ClassQuantMot}

The discussed mechanism of PBHs production requires closed domain walls formation due to the quantum fluctuations of scalar fields at the inflation epoch \cite{2000hep.ph....5271R, 2001JETP...92..921R}. Let us take into account both quantum and classical motion of fields. Consider the scalar field $\Phi$ of mass $m$ and the standard action
\begin{equation}
    \label{action4}
	S = \frac{M_\text{Pl}^{2}}{2}\int \D^{4}x \, \sqrt{g_{4}} \Big[R+
	(\partial\Phi)^2 - m^2 \Phi^2 \Big].
\end{equation}
Here, $M_\text{Pl}$ is the Planck mass. 
The scalar field could be the inflaton field as well as a spectator one.
The field equation in the de Sitter space is represented as \cite{KhlopovRubin}
\begin{gather}
    \frac{\partial \Phi }{\partial t}-\frac{1}{3H}\left[e^{-2Ht}\Delta \Phi - \frac{\partial V(\Phi )}{\partial \Phi}\right] = y(\mathbf{x},t);
    \label{randEq} 
    \\ 
    y(\mathbf{x},t) \equiv \left( -\frac{1}{3H}\frac{\partial ^{2}}{\partial t^{2}}-\frac{\partial }{\partial t}+\frac{1}{3H}e^{-2Ht}\Delta\right) Q( \mathbf{x},t).
    \nonumber
\end{gather}
Here, $Q(x,t)$ is the ``quick'' part of the Fourier  field decomposition. This equation was simplified: we have omitted the second time derivative due to a slow roll approximation and have neglected higher powers of the function $y(\mathbf{x} ,t)$. The latter is supposed to be small so that we may find a solution to the equation in the form
%Eq.\ref{randEq} describes classical motion of the field $\Phi $ under permanent influence of random ``force'' $y$. 
\begin{equation}
    \label{clquant}
    \Phi =\Phi_\text{cl}+\phi.
\end{equation}
The deterministic part of the classical field $\Phi_\text{cl}$ is governed by the equation
\begin{equation}
    \frac{\partial \Phi_\text{cl}}{\partial t} - 
        \frac{1}{3H} \left[ e^{-2Ht}\Delta\Phi_\text{cl} - \frac{\partial V(\Phi_\text{cl})}{\partial \Phi_\text{cl}} \right] = 0,
    \label{fdet}
\end{equation} 
while its random part $\phi$ depends strictly on quantum fluctuations according to the linear equation
\begin{equation}
    \frac{\partial \phi }{\partial t} 
        - \frac{1}{3H} \Big[ e^{-2Ht} \Delta\phi - V^{\prime \prime}(\Phi_\text{cl})\phi \Big] = y(\mathbf{x},t).
    \label{frandom}
\end{equation}
Here, we consider the limit $\Phi_\text{cl} \gg \phi $ which is valid if the random ``force'' $y(\mathbf{x},t)$ is small.

Let us denote
\begin{equation}
    m^2(t) \equiv V^{\prime \prime }(\Phi_\text{cl}).
\end{equation}
The parameter $m$ is positive if we are near the bottom of potential and is imaginary if we are near the potential maximum. It is supposed that $m(t)$ varies slowly during inflation.

We are interested in the super horizon scales where the fluctuations do not depend on the space coordinates. The uniform distribution $\Phi =\Phi (t)$ is governed by the more simple equation
\begin{align}
    & \frac{\partial \Phi_\text{cl} }{\partial t} + \frac{1}{3H}\frac{\partial
    V(\Phi_\text{cl} )}{ \partial \Phi_\text{cl} }=0,  
    \label{lineart} 
    \\ 
    & \frac{\partial \phi }{\partial t}+\mu(t)\phi =y(t);\quad  \mu(t) \equiv \frac{m^{2}(t)}{3H},
    \label{lineart2}
\end{align}
provided that $H(t)=\const$. The correlator of the random function $y(t)$ may be approximated as follows \cite{Rey1987}
\begin{equation}
    \big\langle y(t_{1}) \, y(t_{2}) \big\rangle 
        = D(\mathbf{x}, t_1; \mathbf{x}, t_2)
        = \frac{H}{4\pi ^{2}}\delta(t_{1}-t_{2}).
\end{equation}
The delta function in the rhs of this expression indicates that the random function $y(t)$ is distributed according to the Gauss law with the density
\begin{equation}
    W(y) = \const \, \exp \left[ -\frac 1{2\sigma^2}\int y^2(t) \, \D{t} \right], 
    \quad 
    \sigma =\frac{H^{3/2}}{2\pi}.
\end{equation}

The probability distribution of the function $\phi$ is proportional to that of the function $y(t)$ due to their linear relationship \eqref{lineart2}.
It means that the probability to find the specific value $\phi(t)$ inside some small interval is equal to \cite{feynman2010quantum}
\begin{equation}
    \D{P}(\phi) = \const \, \mathcal{D}\phi \,
    \exp \left[ -\frac{1}{2\sigma^{2}}\int 
    \left[ \frac{\partial \phi }{\partial t}+\mu(t)\phi \right]^{2} \D{t} \right].
\end{equation}

Let's obtain the probability to find a quantum part of the field $\phi_{2}$ at an instant $t_{2}$ provided that a value $\phi _{1}$ at an instant $t_{1}$ is known. Evidently, we have to integrate over all values of the field inside the interval $(t_{1},t_{2})$ and come to the expression
\begin{equation}
    \D{P}(\phi_{2},t_{2};\phi _{1},t_{1}) 
        = \const \, \D\phi_{2}\int\limits_{\phi
    _{1}}^{\phi _{2}}\mathcal{D}\phi \, 
        \exp\left[ -\frac{1}{2\sigma^{2}} \int\limits_{t_{1}}^{t_{2}}
        \left[\frac{\partial \phi}{\partial t}+\mu(t) \phi\right]^{2} \D{t}\right].
    \label{fundistr}
\end{equation} 
The constant factor in this equation is determined by  normalization condition
\begin{equation}
    \int\limits_{-\infty }^{\infty } \diff[]{P(\phi_{2},t_{2}; \phi_{1},t_{1})}{\phi_2} \, \D\phi_{2}
        = 1.
\end{equation}

Functional integral \eqref{fundistr} can be calculated in the standard manner by finding an extreme trajectory of the integral in the exponent 
\begin{equation}
    \ddot{\phi} - \mu^{2}(t)\phi = 0,
           % \exp\bigg[-q^2 \left(\Phi_2 - \Phi_\text{cl}(t_2) - \big(\Phi_1- \Phi_\text{cl}(t_1)\big)e^{-M(t_2)}\right)^2\bigg].
    \label{eq:DDotPhi}
\end{equation}
where the term $\dot{\mu}$ is neglected due to slow variation of $\mu(t)$. The boundary conditions for \eqref{eq:DDotPhi} are as follows
\begin{equation}
    \phi (t_{1})=\phi _{1};
    \quad
    \phi (t_{2})=\phi _{2}.
\end{equation}
Exact solution to this equation is
%(remind that $\mu(t)$ is slowly varying function, in particular, we neglect $\dot{\mu}$.)
\begin{gather}
    \phi(t) = A\exp\big( M(t) \big) + B\exp\big( -M(t) \big);
    \quad 
    M(t)\equiv \int\limits_{t_1}^{t}\mu(t') \, \D{t'};
    \\
    A = \frac{\phi_{2}-\phi_{1}e^{-M(t_2)}}{2\sinh\big(M(t_2)\big)},
    \quad
    B = -\frac{\phi_{2}-\phi_{1}e^{M(t_2)}}{2\sinh\big(M(t_2)\big)}.
\end{gather}
Notice that $M(t_1)=0$ by definition.

Substituting this solution into the integral in the exponent of
the expression \eqref{fundistr} one obtains the desired probability in the saddle point approximation
\begin{gather}
    \label{maindistr} 
    \begin{split}
        \D{P}(\phi_{2}, t_2; \phi_{1}, t_{1}) &= \const \, \D\phi_{2} \, \exp\left[ -\frac{2A^2}{\sigma^2}\int\limits_{t_1}^{t_2} \mu^2(t) e^{2M(t)} \, \D{t} \right]= \\
        & =  \const \, \D\phi_{2} \, \exp\bigg[-q^2 \Big(\phi_2-\phi_1 e^{-M(t_2)}\Big)^2\bigg], 
    \end{split}
    \\
    q^2\equiv \frac{1}{2\sigma^2\sinh^2\big(M(t_2)\big)}\int\limits_{t_1}^{t_2} \mu^2(t)e^{2M(t)} \, \D{t}.
\end{gather}
It describes the probability to find specific value of "quantum" part of the field \eqref{clquant}. The "classical" part of the field $\Phi_\text{cl}$ is incorporated into the function $M(t)$.
The probability for the field value $\Phi$ (the distribution function $f$) is easily obtained by substitution $\phi(t)=\Phi(t)-\Phi_\text{cl}(t)$ into the formula above.
\begin{multline}
    f(\Phi_2, t) = 
    \diff[]{P(\Phi_2, t; \Phi_1, t_1)}{\Phi_2} =
    %\D{P}(\Phi_2, t; \Phi_1, t_1)             % = \D\Phi_2 
    \\
    = \sqrt{\cfrac{q^2}{\pi}} 
        \exp\bigg[-q^2 \left(\Phi_2 - \Phi_\text{cl}(t_2) - \big(\Phi_1- \Phi_\text{cl}(t_1)\big)e^{-M(t_2)}\right)^2\bigg].
    \label{dP}
\end{multline}
%Here, $\phi_i= \Phi_i - \Phi_\text{cl}(t_i)$, $i=1,2$.
The limit $m\to 0$ restores the textbook formula.

The aim of the next section is to demonstrate how the obtained formulas can be applied to a particular scalar field potential. 
It is assumed that the potential may possess many extremes of different kind. 
In our consideration, we choose a part of phase space containing two maxima and at least one saddle point.

\section{The PBH formation}
\label{sec:PBHForm}

In this section, we show that the classical motion of fields together with their quantum fluctuations influence the PBH mass spectra. To this end, we have to find the classical trajectory and use the probability \eqref{dP} derived above. 

The fields move between potential local maxima that leads to complicated spectra of fluctuations. The latter are discussed in papers \cite{2007LNP...738..275W, 2012JCAP...08..022K}. At the same time, the presence of saddle points is the reason of the closed domain walls formation, see details in \cite{2018JCAP...04..042G, 2021Physics.3.563}. In the following, they could collapse to black holes \cite{2001JETP...92..921R, 2019EPJC...79..246B}.

Let us consider the model of two real scalar fields with the Lagrangian
\begin{equation}
    \mathcal{L} = \cfrac{1}{2}\,\big(\partial_\mu\Phi \partial^\mu\Phi + \partial_\mu\Chi \partial^\mu\Chi\big) - V(\Phi, \Chi).
    \label{eq:L}
\end{equation}
We choose the potential possessing $n$ peaks and saddle points
\begin{equation}
    \begin{gathered}
        V(\Phi, \Chi) 
        = \cfrac{m_\phi^2}{2} \, \Phi^2 + \cfrac{m_\chi^2}{2} \, \Chi^2 + \sum_{i=1}^n \delta V_i(\Phi, \Chi),
        \\
        \delta V_i(\Phi, \Chi) 
        = \Lambda_{i}^4 \exp
        \left( 
            -\left(\frac{\Phi-\phi_{i}}{\Delta_i}\right)^2 
            -\left(\frac{\Chi-\chi_{i}}{\Delta_i}\right)^2 
        \right).
    \end{gathered}
    \label{eq:V}
\end{equation}
Here, $\delta V_i$ describes the $i$-th local maximum. The global minimum of the potential is located at the point $(\phi_{\min}, \chi_{\min}) = (0,0)$ with exponentially small errors. Hereinafter, all variables are taken in the Hubble units $H$ where $H\approx10^{13}$~GeV at the inflationary epoch. 

For our estimates, we choose the fields masses $m_\phi = 0.4$ and $m_\chi = 0.5$. For simplicity, we consider the potential with two peaks ($n=2$) with the coordinates $\phi_1 = -9.0$, $\chi_1 = 3.0$ and $\phi_2 = -1.7$, $\chi_2 = 4.5$. The parameters corresponding to the peaks heights are $\Lambda_1 = 3.0$ and $\Lambda_2 = 1.5$, and the peaks widths are set with $\Delta_1 = 0.5$ and $\Delta_2 = 1.5$. The initial fields values are $\phi_\text{in}=-8.0$ and $\chi_\text{in}=45.0$. Note, all chosen parameters have the values $\sim\mathcal{O}(1)$.

Following Section~\ref{sec:ClassQuantMot}, the first step consists of finding the classical trajectory $\Phi_{\text{cl}}(t)$, $\Chi_{\text{cl}}(t)$ of the fields $\Phi$, $\Chi$.
Starting from the initial values $(\phi_\text{in},\chi_\text{in})$ at the inflation epoch, the scalar fields tend to the potential minimum. The process is described by the classical motion equations
\begin{equation}
    \begin{cases}
        \Phi_{\text{cl},tt} + 3H \, \Phi_{\text{cl},t} 
            + \diffp{V(\Phi_{\text{cl}}, \Chi_{\text{cl}})}{{\Phi_{\text{cl}}}} = 0,
        \\
        \Chi_{\text{cl}, tt} + 3H \, \Chi_{\text{cl},t} 
            + \diffp{V(\Phi_{\text{cl}}, \Chi_{\text{cl}})}{{\Chi_\text{cl}}} = 0.
    \end{cases}
    \label{eq:ClassDynam}
\end{equation}

\begin{figure}[t]
    \centering
    \includegraphics[width = 0.8\linewidth]{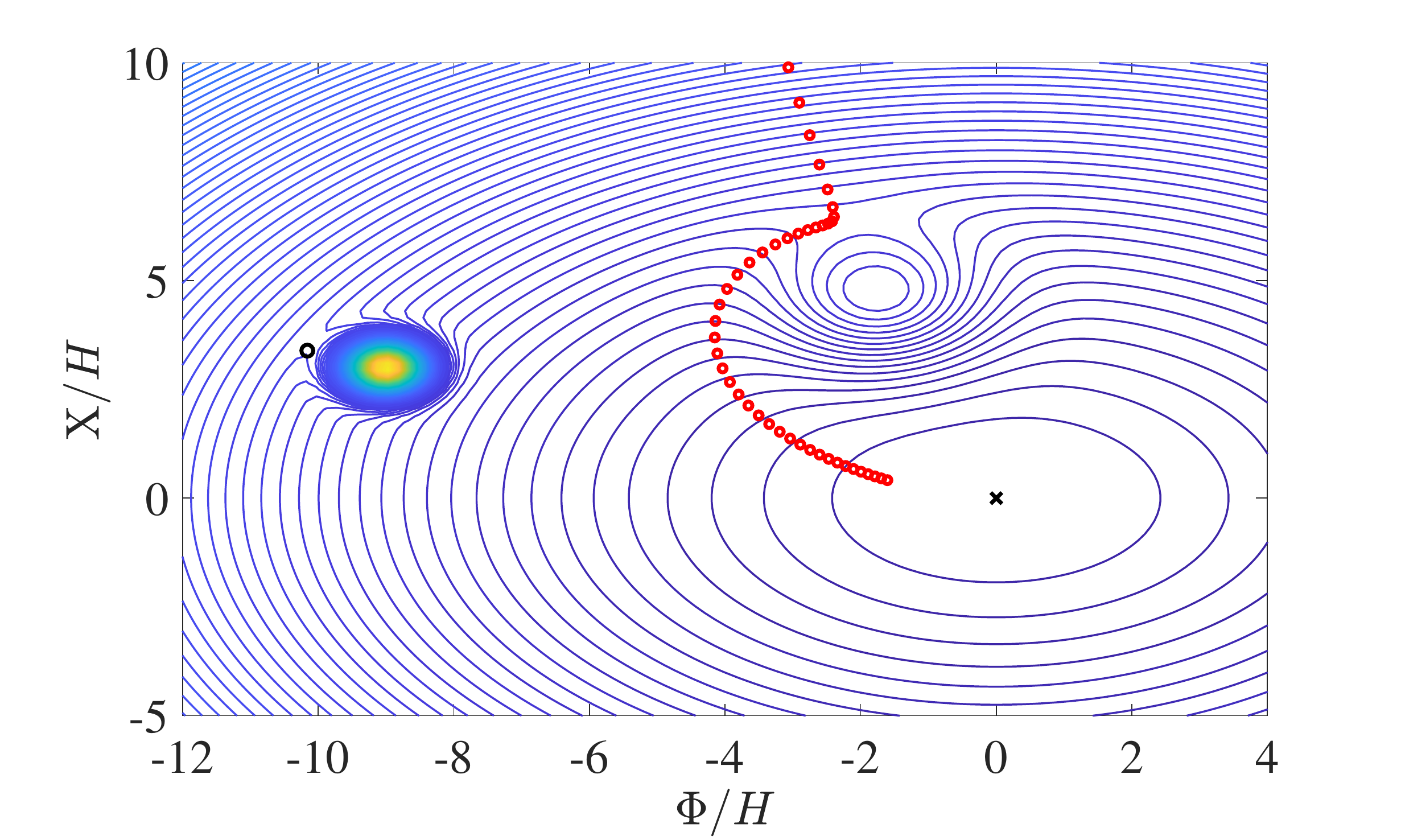}
    \caption{The contour plot of the potential \eqref{eq:V} with the parameters $m_1 = 0.4$, $m_2 = 0.5$, $\Lambda_1 = 3.0$, $\Lambda_2 = 1.5$, $\phi_1 = -9.0$, $\chi_1 = 3.0$, $\phi_2 = -1.7$, $\chi_2 = 4.5$, $\Delta_1 = 0.5$, $\Delta_2 = 1.5$ is shown. The red circles illustrate the classical trajectory of the fields $\Phi$, $\Chi$, and the black cross shows the potential minimum. The initial fields values for \eqref{eq:ClassDynam} are $\phi_\text{in}=-8.0$ and $\chi_\text{in}=45.0$. }
    \label{fig:ClassDynam}
\end{figure}

The classical evolution of the fields and the form of the specific potential are shown in Figure~\ref{fig:ClassDynam}. 
At the same time, quantum fluctuations lead to fields ``diffusion'' during inflation. 
%In the flat potential approximation, quantum fluctuations may be described by the Fokker-Plank equation giving the probability density function $f(t, \phi,\chi)$ to find fields $(\phi,\chi)$ in any point of the physical space. We assume that the fluctuations of fields $\phi$ and $\chi$  are independent. Thus, the probability density function is separated over the  fields $\phi$ and $\chi$????. 
The probability density $f$ to find the fields $\Phi$ or $\Chi$ in any point of the physical space is given by formula \eqref{dP}.

Both $\Phi$ and $\Chi$ distributions depend on a classical position of the fields at the instant $t$. In our estimates, we suppose that the probability function for the quantum parts $\phi$ or $\chi$ of the fields is separated into two independent fluctuation processes
\begin{equation}
    f(t, \phi, \chi) = f_\phi(t, \phi) \, f_\chi(t, \chi) = \diff[]{P}{\phi}\diff[]{P}{\chi},
    \label{eq:f}
\end{equation}
where distribution functions of each field $\phi, \chi$ can be defined as in \eqref{maindistr}.

As noted above, the fields should reach a saddle point of potential for domain wall formation. Suppose that some quantum fluctuation crosses a saddle point and ends up at a point of the area $\Omega$. As was shown in \cite{2018JCAP...04..042G, 2021Physics.3.563}, it causes nontrivial field solutions of the system \eqref{eq:ClassDynam} characterized by a nonzero winding number. Such configurations might lead to the domain walls formation after the inflation is finished. Detailed explanation might be found in \cite{2018JCAP...04..042G}. The calculation of an exact shape of the area $\Omega$ is a separate, quite complicated task, so that we limit ourselves with the following approximation. Let us assume that the area $\Omega$ is bordered by two lines $\chi_\text{cl-sp}(\Phi)$ and $\chi_\text{min-sp}(\Phi)$ in the phase space. The first line connects the classical value at the instant $t$ and the saddle point $(\phi_\text{sp}, \chi_\text{sp})$
\begin{equation}
    \chi_\text{cl-sp}(\Phi) = \chi_\text{sp} + 
        \cfrac{\Phi - \phi_\text{sp}}{\phi_\text{cl}(t) - \phi_\text{sp}}\,
        (\chi_\text{cl}(t) - \phi_\text{sp}).
\end{equation}
The second one connects the vacuum value and the saddle point
\begin{equation}
    \chi_\text{min-sp}(\Phi) = \chi_\text{sp} + 
        \cfrac{\Phi - \phi_\text{sp}}{\phi_\text{min} - \phi_\text{sp}}\,
        (\chi_\text{min} - \phi_\text{sp}).
\end{equation}
% \textcolor{red}{Above are Big initial latters!!}
Thus, the probability for the fields to attain the area $\Omega$ where domain walls might form is calculated by integrating \eqref{eq:f}
\begin{equation}
    P(t) = \iint\limits_{\Omega} f(\phi, \chi, t) \, \D \Chi \, \D \Phi \,
         = \int\limits_{\phi_\text{sp}}^{+\infty}  f_\phi(\phi, t) \, \D \Phi
           \int\limits_{\chi_\text{cl-sp}(\Phi)}^{\chi_\text{min-sp}(\Phi)} f_\chi(\chi, t) \, \D \Chi.
    \label{eq:Prob}
\end{equation}
Here, both distribution functions $f_\phi$ and $f_\chi$ are defined in \eqref{dP} and according to \eqref{clquant} $\phi=\Phi - \Phi_{\text{cl}},\, \chi=\Chi - \Chi_{\text{cl}}$. The algorithm for calculating the probability \eqref{eq:Prob} discussed above is more accurate then that used in the previous papers. 

Now, let us find the mass spectra of primordial black holes. In the considered model, they are formed due to collapse of domain walls. Here, we briefly reproduce the idea, while details may be found in the review \cite{2019EPJC...79..246B}. 
% As was shown in \cite{2018JCAP...04..042G}, field configurations with nonzero winding number are formed due to the quantum fluctuations.
As was shown in \cite{2018JCAP...04..042G, 2021Physics.3.563}, domain walls might be formed due to the quantum fluctuations in field models with potential possessing at least one saddle point and a local maximum. The protosoliton is formed if the fields achieve a saddle point (in our model, we have noted this area as $\Omega$). These protosoliton field configurations are quickly expanded during inflation. The final scale of such configuration depends on a e-fold number $N$. More definitely, the configuration scale is stretched in the factor $\sim e^{N_\text{inf}-N}$ to the end of inflation. The soliton is quickly formed after the end of inflation. The total mass of the field configuration is proportional to its area. Evidently, it could collapse into a black hole after the end of inflation \cite{2000hep.ph....5271R, 2001JETP...92..921R, 2002astro.ph..2505K}.

The regions number where the fields reach the critical values $\phi_\text{cr}$ and $\chi_\text{cr}$ belonging to $\Omega$ can be found as
\begin{equation}
    n(t) = P(t) \, e^{3Ht}.
    \label{eq:n(t)}
\end{equation}
Here, the term $e^{3Ht}$ is the number of causally independent regions of the size $H^{-1}$ at the instant $t$ from the beginning of inflation. After the end of inflation at $t=N_\text{inf}H^{-1}=60H^{-1}$, the size of each region is expanded
\begin{equation}
    r_0(t) = H^{-1} \exp \big( N_\text{inf} - H t \big) \gg r_{h,0}.
    \label{eq:r_inf}
\end{equation}
Here, $N_\text{inf}\approx60$ is the total e-folds number, and $r_{h,0}$ is the horizon size at the end of inflation. At the radiation stage (RD), each region expands as $\propto \sqrt{\tau}$ while the horizon size is $r_h = H^{-1}(\tau) = 2\tau$. Here, $\tau$ is time after the beginning of the RD epoch. After a domain wall goes under the horizon, its collapse begins (the details of the process taking into account detachment from the Hubble flow may be found in \cite{2005APh....23..265K, 2008ARep...52..779D, 2019EPJC...79..246B}). Thus, the maximal size of a domain wall can be written as the function of the instant $t$
\begin{equation}
    r(t) \approx \cfrac{1}{2} \cfrac{e^{2(N_\text{inf} - Ht)}}{HN_\text{inf}}.
    \label{eq:r(t)}
\end{equation}
%\textbf{This equation allows to get the dependence $t(r)$. After its substitution into \eqref{eq:n(t)} - UNCLEAR!}, 
After eliminating of $t$ from \eqref{eq:n(t)} and \eqref{eq:r(t)}, one can finally get the distribution $n(r)$ of closed walls sizes which can be rearranged into the mass spectrum.

\begin{figure}[t]
    \centering
    \includegraphics[width = 0.7\linewidth]{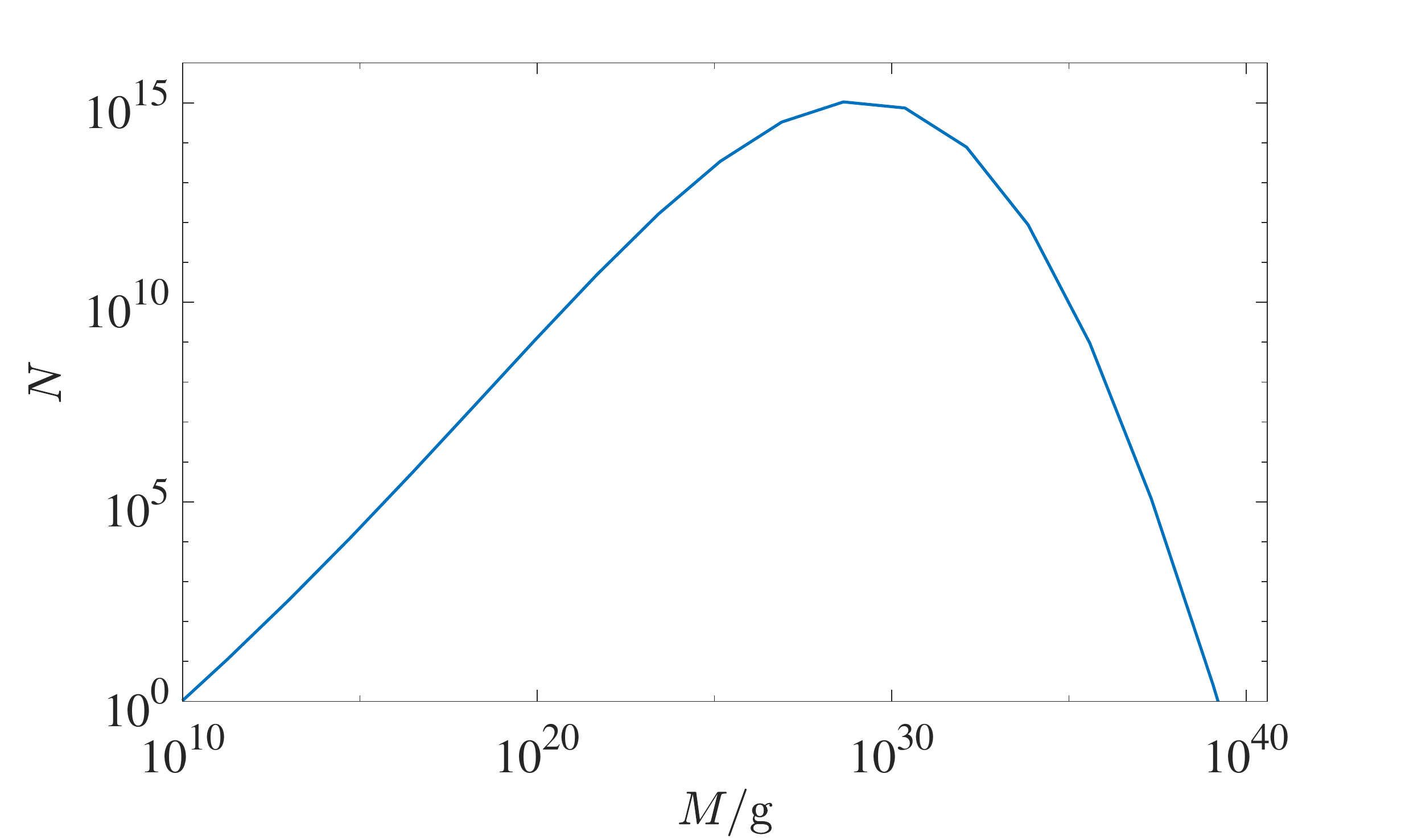}
    \caption{The PBH mass distribution is shown.}
    \label{fig:MassSpectrum}
\end{figure}

Next, we have to find masses of PBHs. For simplicity, we assume total energy of domain wall converts to a black hole mass during its collapse and neglect nonsphericity of a domain wall and losses caused by gravitational waves. The energy density of a domain wall might be found by a common way. The energy momentum tensor for the Lagrangian \eqref{eq:L} is given by
\begin{equation}
    \begin{split}
        T^\mu_\nu %&= \sum_{i=1,2} 
            %\diffp{\mathcal{L}}{{\big(\partial_\mu \varphi_i\big)}} \, \partial_\nu\varphi_i - \mathcal{L} \, \delta^\mu_\nu 
            %= \\
            &= \sum_{i=1,2}  \bigg( \partial^\mu \varphi_i \, \partial_\nu \varphi_i 
            - \cfrac{1}{2} \, \partial^\alpha \varphi_i \, \partial_\alpha \varphi_i \delta^\mu_\nu \bigg) + V \, \delta^\mu_\nu,
    \end{split}
\end{equation}
where $\varphi_1$, $\varphi_2$ correspond to the fields $\Phi_\text{cl}$ and $\Chi_\text{cl}$, respectively. Then, the energy density of a domain wall is found to be
\begin{equation}
    \varepsilon(x) = T^0_0 = \cfrac{1}{2} \, \sum_{i=1,2} 
        \big( (\partial_t \varphi_i)^2 + (\partial_x \varphi_i)^2\big) + V.
    \label{eq:eps}
\end{equation}
Upon integrating \eqref{eq:eps} over the all possible values of $x$ (infinite interval), the surface energy density of a domain wall $\sigma$ may be found. 
% we get
% \begin{equation}
%     \sigma = \int\limits_{-\infty}^{+\infty} \varepsilon(x) \,\D x.
% \end{equation}
Finally, masses of black holes are $M\big(r(t)\big) \simeq 4 \pi \sigma r^2(t)$. Taking into account \eqref{eq:n(t)} and \eqref{eq:r(t)}, one can find the mass spectrum of primordial black holes. 
The PBH mass distribution for the parameters of the Lagrangian \eqref{eq:L} is shown in Figure~\ref{fig:MassSpectrum}. 
Note, the mass spectrum has the non-power form due to taking into account both the quantum and classical motion of scalar fields and the compound form of the potential leading to nontrivial classical fields trajectory.
The obtained spectrum is free from overproduction of light PBHs and, therefore, is much more adaptable to new observational effects. 
We leave the detailed analysis of these possibilities for future research.

\section{Conclusion}
\label{sec:Concl}

In this note, we have shown that the mechanism of PBHs formation in the second order phase transitions of scalar fields might produce a wide variety of the PBH mass spectra. It is expected that observations will help to select an appropriate one.
The key point is the classical field motion which was taken into account together with the quantum fluctuations at the inflationary stage. 
We show here that the probability to find a particular value of scalar field at a space point depends on its classical dynamics. 
We have derived the appropriate analytical formula and have applied it to obtain one of the PBH mass spectra.
The elaborated method is the useful tool to fit an observable spectrum in the near future.

\section*{Acknowledgements}
This research was funded by the Ministry of Science and Higher Education of the Russian
Federation, Project ``Fundamental properties of elementary particles and cosmology'' №~0723-2020-0041.

\printbibliography

\end{document}